\documentstyle[aps,multicol,tabularx,rotate,epsf]{revtex}
\begin{document}
\title{Forces Induced by Non-Equilibrium Fluctuations: The Soret-Casimir Effect}
\author{Ali Najafi and Ramin Golestanian}
\address{Institute for Advanced Studies in Basic Sciences,
Zanjan 45195-159, Iran \\}
\date{\today}
\maketitle

\begin{abstract}

The notion of fluctuation--induced forces is generalized to the
cases where the fluctuations have nonequilibrium origin. It is
shown that a net force is exerted on a single flat plate that
restricts scale-free fluctuations of a scalar field in a
temperature gradient. This force tends to push the object to the
colder regions, which is a manifestation of thermophoresis or the
Soret effect. In the classic two-plate geometry, it is shown that
the Casimir forces exerted on the two plates differ from each
other, and thus the Newton's third law is violated.
\medskip

\noindent Pacs numbers: 05.40.-a, 82.70.-y, 05.70.Jk, 82.70.Dd,
05.70.Np

\end{abstract}

\begin{multicols}{2}

\section{Introduction}  \label{sec:Intro}

In his pioneering work in 1948, Casimir introduced the notion of
fluctuation--induced forces when he showed that quantum
fluctuations of the electromagnetic fields can generate measurable
long-ranged forces between conducting plates \cite{Casimir}. The
idea, however, has been subsequently generalized to various cases
where: (i) the fluctuations are of classical thermal origin, (ii)
the fluctuating media are complex fluids, and (iii) the boundaries
are moving \cite{Dzya,mostepa,Krech,KGRMP}.

In light of the extensive development of the original idea of
fluctuation--induced interactions, it may seem quite natural to
ask what happens if the fluctuations that cause the interaction
are of nonequilibrium nature. A first attempt towards answering
this question was made by G.I. Taylor already in 1951, when he was
studying the swimming mechanism of flagella (tails of spermatozoa)
and the theoretically challenging question of the possibility of
``swimming at low Reynolds number'' \cite{taylor}. In this classic
work, Taylor has shown that two flagella undergoing wavelike
harmonic deformations have a tendency to attract each other,
because their undulations introduce (nonequilibrium) hydrodynamic
fluctuations in the surrounding viscous fluid that could induce an
effective interaction between them. More recently, another
generalization has been introduced where localized sources drive
the fluctuating medium out of equilibrium \cite{bartolo}. We also
note that the idea of depletion forces \cite{osawa}, which are
also fluctuation--induced in origin, has been generalized to
systems driven out of equilibrium \cite{depletion}.

A particularly interesting way of driving a thermally fluctuating
system out of equilibrium is by imposing spatial or temporal
variations in the temperature profile. Many interesting phenomena
are known to appear under such conditions, ranging from anomaly in
diffusion to the old problem of thermophoresis or Soret effect
\cite{temgrad,soret}. In this effect, whose microscopic mechanism
is still subject to debate \cite{soret3}, solutions of colloidal
particles \cite{therdiff} or polymers \cite{soret2} that are
placed in a temperature gradient experience separation, which
signals the appearance of a net driving force on particles as a
result of the temperature gradient. It has also been recently
shown that inhomogeneity in the temperature profile could lead to
deviation of the self-diffusion coefficient of a tracer undergoing
Brownian motion from the simplistic local Einstein's relation
\cite{tracer}.

\begin{figure}
\centerline{ \epsfxsize=6truecm \epsffile{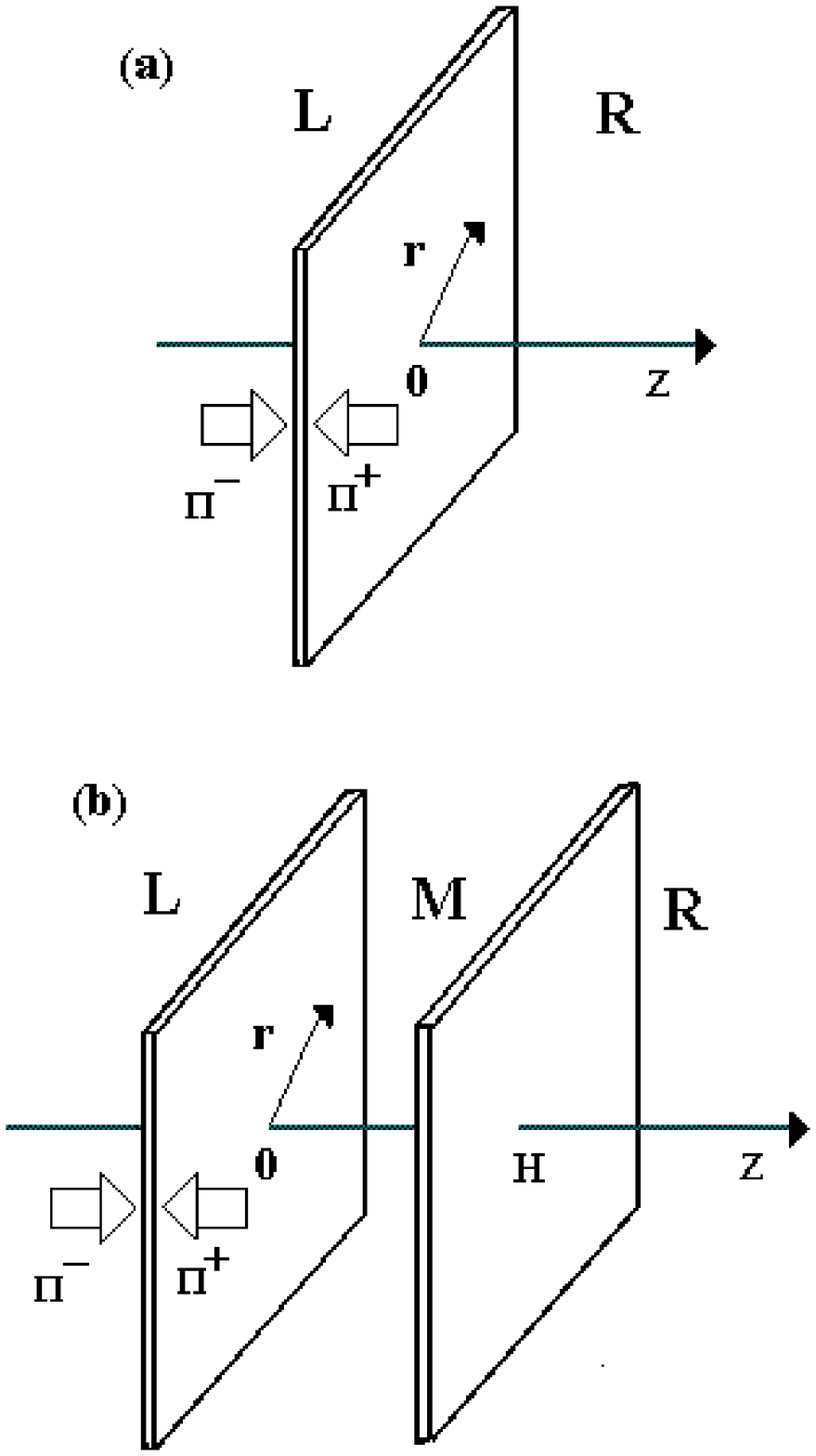} }
\vskip0.9truecm \caption{(a) A single plate immersed in a
fluctuating medium. (b) Two parallel plates separated by a
distance $H$ in a same medium. In both cases $\Pi^{\pm}$ represent
the stresses exerted on a plate from the right- or the left-hand
side of it. \hskip12truecm } \label{onetwo}
\end{figure}

Here, we attempt to generalize the notion of Casimir forces for
nonequilibrium systems in which temperature profile is not uniform
\cite{Schmitz}. We consider a near-equilibrium fluctuating system
in which we can define local and instantaneous temperature. We
study the dynamics of a classical scalar field governed by a
Langevin equation, in which the strength of the noise is
proportional to the local and instantaneous temperature in the
system. We consider external objects that restrict the
fluctuations of the scalar field and calculate the resulting
fluctuation--induced stress on these objects. We concentrate on
the examples of a single flat plate as well as two parallel
plates, as depicted in Fig. \ref{onetwo}, immersed in various
temperature profiles. For the case of a single plate, we find that
a net force is exerted on the plate due to the imbalance of the
fluctuation--induced forces from the two sides, which has a
tendency to push it to the colder regions. We also calculate the
Casimir force between two parallel plates in nonuniform
temperature profiles and find that the two plates experience
different forces in violation of the third law of Newton.

The rest of the paper is organized as follows: In Secs.
\ref{sec:Model} and \ref{sec:relaxation}, we introduce the
Langevin dynamics of the scalar field, which we expect to capture
the most physically relevant consequences of nonequilibrium
temperature profiles, and define the quantities of interest that
are to be calculated. Section \ref{sec:single} is devoted to the
discussion of the example of a single plate placed in various
temperature profiles, which is followed by the corresponding
discussions for the two-plate geometry in Sec. \ref{sec:twoplate}.
Finally Sec. \ref{sec:conc} concludes the paper, while an Appendix
outlines some details of the calculations.

\section{The Model}     \label{sec:Model}

Let us consider an equilibrium system described by a scalar field
$\phi({\bf R},t)$ that undergoes thermal fluctuations in a space
that is bounded by a number of external objects. The field could
represent a component of the electromagnetic field (e.g. the
electric potential) in a dielectric medium or vacuum
\cite{Casimir,Dzya,mostepa}, an order parameter field for a
critical binary mixture or a magnetic system \cite{Krech}, a
massless Goldstone mode arising from a continuous symmetry
breaking such as nematic liquid crystals or superfluid helium, an
elastic deformation field for fluctuating membranes and surfaces,
or the electrostatic potential in charged fluids at very low salt
concentrations \cite{KGRMP}. In all of the above systems, it is
possible to write down a Hamiltonian for the fluctuations around
the equilibrium state, which in Gaussian approximation reads
\begin{equation}
{\cal H}=\frac{K}{2}\int d^3{\bf R} \; \left[\nabla\phi({\bf R})
\right]^2, \label{hamiltonian}
\end{equation}
where $K$ is an elastic modulus describing the stiffness of the
system for the fluctuations around equilibrium state.

We consider the situation where temperature has a slowly varying
profile such that thermal equilibrium can be achieved locally.
This means that we have many heat reservoirs locally in contact
with our system, which have different temperatures. The
temperature difference between neighboring reservoirs should be
sufficiently small so that the condition of local equilibrium is
fulfilled. Similarly, we can consider a temperature profile with
temporal variations that are sufficiently slow such that
instantaneous equilibrium state can be defined. We would like to
drive the system described by the Hamiltonian in Eq.
(\ref{hamiltonian}) out of equilibrium by imposing such
temperature profiles and examine the corresponding
fluctuation--induced forces exerted on external boundaries in the
medium.

Our specific choices of boundaries are sketched in
Fig.~\ref{onetwo}. We consider both the case of a single plate of
area $A$, as well as two such parallel plates separated by a
distance $H$. We choose the $z$ axis to be the normal direction to
the plates, such that the three dimensional position vector can be
represented as ${\bf R}=({\bf r},z)$. We assume the Dirichlet
boundary condition, which means that the fluctuating field $\phi$
is restricted to vanish on the plates.

The physical quantity that we would like to calculate for such
systems is the pressure or the normal force per unit area exerted
on each plate from the fluctuating medium. In the prescribed
geometry, this will be $\Pi_{zz}$---the $zz$ component of the
stress tensor. Using the Hamiltonian in Eq. (\ref{hamiltonian}),
one can calculate the components of the stress tensor in terms of
the two point correlation functions of the field. For example, the
local pressure exerted from the right side on the plate that is
located at $z=0$ is given as
\begin{equation}
\Pi_{zz}^{+}({\bf
r},t)=-\frac{K}{2}\partial_z\partial_{z'}\left\langle\phi({\bf
r},z,t)\phi({\bf r},z',t)\right\rangle \mid_{z,z'\rightarrow
0^+},\label{pi-def}
\end{equation}
where the minus sign denotes that this pressure tends to push the
plate to the $-z$ direction. (Note that the normal to plate at
this point is in $+z$ direction.) A similar definition and
expression can be given for the pressure from the left side
$\Pi_{zz}^{-}$, and the total normal force per unit area
\begin{equation}
\Pi_{\rm tot}=\Pi_{zz}^{-}-\Pi_{zz}^{+},\label{pi-tot-def}
\end{equation}
exerted on the plate in the positive $z$ direction. Note that due
to the translational symmetry of the infinite plates in the
parallel directions, the lateral forces are all zero, even for the
case where temperature is not uniform.

\section{Dynamical relaxation to local equilibrium}    \label{sec:relaxation}

For calculating the correlation functions in the local and
instantaneous near-equilibrium states resulted from a prescribed
temperature profile, we use the fact that the such states could be
reachable through an equilibrium dynamical relaxation. We assume a
dynamical Langevin equation as
\begin{equation}
\gamma\partial_{t}\phi({\bf R},t)= K \nabla^2\phi({\bf
R},t)+\eta({\bf R},t), \label{langevin}
\end{equation}
with the boundary conditions $\phi({\bf r},0,t)=\phi({\bf
r},H,t)=0$. Here $\gamma$ is the friction coefficient of the
system, and the random force $\eta({\bf r,t})$ is a Langevin noise
whose spectrum is given by the local and instantaneous
fluctuation---dissipation theorem:
\begin{equation}
\left\langle \eta({\bf R},t) \eta({\bf R}',t')\right\rangle=2
\gamma k_{\rm B} T({\bf R},t) \delta^3({\bf R}-{\bf R}')
\delta(t-t'), \label{noise}
\end{equation}
where $T({\bf R},t)$ is the temperature profile in the system.

For solving the above Langevin equation, we define the appropriate
Green's function as the solution of the following equation
\begin{equation}
(\gamma \partial_{t} -K\nabla^2)G({\bf R},t;{\bf
R}',t')=\delta^3({\bf R}-{\bf R}')\delta(t-t'), \label{green}
\end{equation}
with the boundary conditions
\begin{equation}
G({\bf R},t;{\bf R}',t')|_{{\bf R}~{\rm on~
boundary}}=0,\label{BC-on-G}
\end{equation}
for each value of ${\bf R}'$. The solution of the Langevin
equation can then be written as
\begin{equation}
\phi({\bf R},t)=\int d^3{\bf R}' d t' \; G({\bf R},t;{\bf R}',t')
\; \eta({\bf R}',t'),\label{phi-G-eta}
\end{equation}
which yields the two point correlation function of the field as
\begin{eqnarray}
\left\langle \phi({\bf R},t) \phi({\bf R}',t')\right\rangle&=& 2
\gamma k_{\rm B} \int d^3{\bf R}_1 d t_1 \; T({\bf R}_1,t_1)
\nonumber \\
&& \times \; G({\bf R},t;{\bf R}_1,t_1) \; G({\bf R}',t';{\bf
R}_1,t_1).\label{phi-phi-corr}
\end{eqnarray}
Note that the fluctuations in the $\phi$-field are affected by the
entire temperature profile through the scale-free relaxation
dynamics governed by Eq. (\ref{langevin}).

To obtain the form of the Green's function for the specific
geometries discussed above, we proceed by direct solution of the
differential equation. We focus only on the two-plate case, since
the single-plate Green's function can be deduced from it by simply
setting $H=0$. Because of the translational symmetry in time, as
well as the space direction parallel to the plates, we introduce
the Fourier transformation
\begin{equation}
G_{{\bf q},\omega}(z,z')=\int d^2{\bf r} d t \; e^{i {\bf
q}\cdot({\bf r}-{\bf r}^{'})-i \omega (t-t')} \; G({\bf
r},z,t;{\bf r}',z',t'),\label{fourier-def}
\end{equation}
which satisfies the following equation:
\begin{equation}
(Q_{-}^{2}-\partial_{z}^{2})G_{{\bf
q},\omega}(z,z')=\frac{1}{K}\delta(z-z'), \label{GF}
\end{equation}
where $Q_{\pm}=\sqrt{q^2 \pm i \gamma \omega/K}$. We can devide
the space into three distinct subspaces: left (L), middle (M), and
right (R), as shown in Fig. 1(b), and label the Green's function
in each subspace with the corresponding indices: L, M, or R. The
boundary conditions can be summarized as
\begin{eqnarray}
G^{L,M}_{{\bf q},\omega}(0,z')&=& G^{M,R}_{{\bf q},\omega}(L,z')=0, \nonumber\\
G^{L}_{{\bf q},\omega}(-\infty,z')&=& G^{R}_{{\bf
q},\omega}(\infty,z')=0, \nonumber \\
G^{L,M,R}_{{\bf q},\omega}(z'^{-},z')&=&G^{L,M,R}_{{\bf q},\omega}(z'^{+},z'), \nonumber\\
\partial_z G^{L,M,R}_{{\bf q},\omega}(z'^{-},z')
&-& \partial_z G^{L,M,R}_{{\bf
q},\omega}(z'^{+},z'))=\frac{1}{K}.\label{BC-GGGGG}
\end{eqnarray}
Using the above boundary conditions we can solve Eq.(\ref{GF}) and
directly obtain the closed form for Green's function in the
different parts of the space as
\begin{eqnarray}
G_{{\bf q},\omega}^{L}(z,z')&=&\left\{
\begin{array}{l}
-\sinh [Q_{-}z] \; \frac{\exp[{Q_{-} z'}]}{K Q_{-}}~;~z>z', \nonumber\\
\\
-\sinh [Q_{-}z']\; \frac{\exp[{Q_{-} z}]}{K Q_{-}}~;~z<z', \nonumber\\
\end{array}
\right.\\ \nonumber\\
G_{{\bf q},\omega}^{M}(z,z')&=&\left\{
  \begin{array}{l}
   -\frac{\sinh [Q_{-}z']\; \sinh [Q_{-}(z-H)]}{K Q_{-} \; \sinh [Q_{-}H]}~;~z>z', \nonumber\\
   \\
   -\frac{\sinh [Q_{-}z] \; \sinh [Q_{-}(z'-H)]}{K Q_{-} \; \sinh [Q_{-}H]}~;~ z<z', \nonumber\\
  \end{array}
                   \right.\\ \nonumber\\
G_{{\bf q},\omega}^{R}(z,z')&=&\left\{
  \begin{array}{l}
    \sinh [Q_{-}(z'-H)] \; \frac{\exp[{-Q_{-}(z-H)}]}{K Q_{-}}~;~z>z', \nonumber\\
    \\
    \sinh [Q_{-}(z-H)] \; \frac{\exp[{-Q_{-}(z'-H)}]}{K Q_{-}}~;~ z<z', \nonumber\\
  \end{array}
                   \right.\\
                   \label{G-solution}
\end{eqnarray}

The above expressions for the Green's function can be used to
calculate the normal stress exerted on the plates for arbitrarily
given temperature profiles. While the equilibrium
fluctuation--induced forces are universal, it appears that forces
induced by nonequilibrium fluctuations are not universal and do
depend on the specific choice of dynamics as well as the
microscopic details, as will be shown in the next two sections.

\section{Single-Plate Geometry} \label{sec:single}

In this section we focus on the example of a single plate that is
embedded in a fluctuating background with local temperature. When
the temperature profile is uniform, the stress resulted from the
field configurations on the left side cancels completely with the
corresponding counterpart coming from the right side, and the
plate experiences no net force. However, when temperature is not
uniform the strength of the fluctuations are also not uniform, and
we expect a force imbalance due to the asymmetry caused by the
temperature gradient. To further simplify the problem, we assume
that temperature only depends on the $z$ coordinate, i.e. it is
independent of time and the parallel coordinates. The two
contributions to the normal stress at the position of the plate
can be written as
\begin{equation}
\Pi^-=-\frac{\gamma k_{\rm B}}{K} \int\frac{d^2{\bf
q}d\omega}{(2\pi)^3} \int_{-\infty}^{0}d z \; T(z) \;
e^{(Q_{+}+Q_{-})z},\label{1P-Pi-}
\end{equation}
and
\begin{equation}
\Pi^+=-\frac{\gamma k_{\rm B}}{K} \int\frac{d^2{\bf
q}d\omega}{(2\pi)^3} \int_{0}^{\infty} d z \; T(z) \;
e^{-(Q_{+}+Q_{-})z}.\label{1P-Pi+}
\end{equation}
Due to the nonuniformity of the temperature profile, the
nonuniversal contributions to the stress on the two sides do not
cancel out. Therefore, the forces mediated by nonequilibrium
fluctuations are not universal and they also depend on the
specific imposed temperature profile. Examining the behavior of
Eqs. (\ref{1P-Pi-}) and (\ref{1P-Pi+}) above, we can distinguish
between three different categories: (i) the case where temperature
is continuous on the plate and its asymptotic values are equal at
$z=-\infty$ and $z=+\infty$, (ii) the case where temperature is
continuous on the plate but its asymptotic values at the two
infinities are not the same, and (iii) the case where the
temperature is not continuous on the plate. Each category is
examined separately below.

\begin{figure}
\centerline{ \epsfxsize=8truecm \epsffile{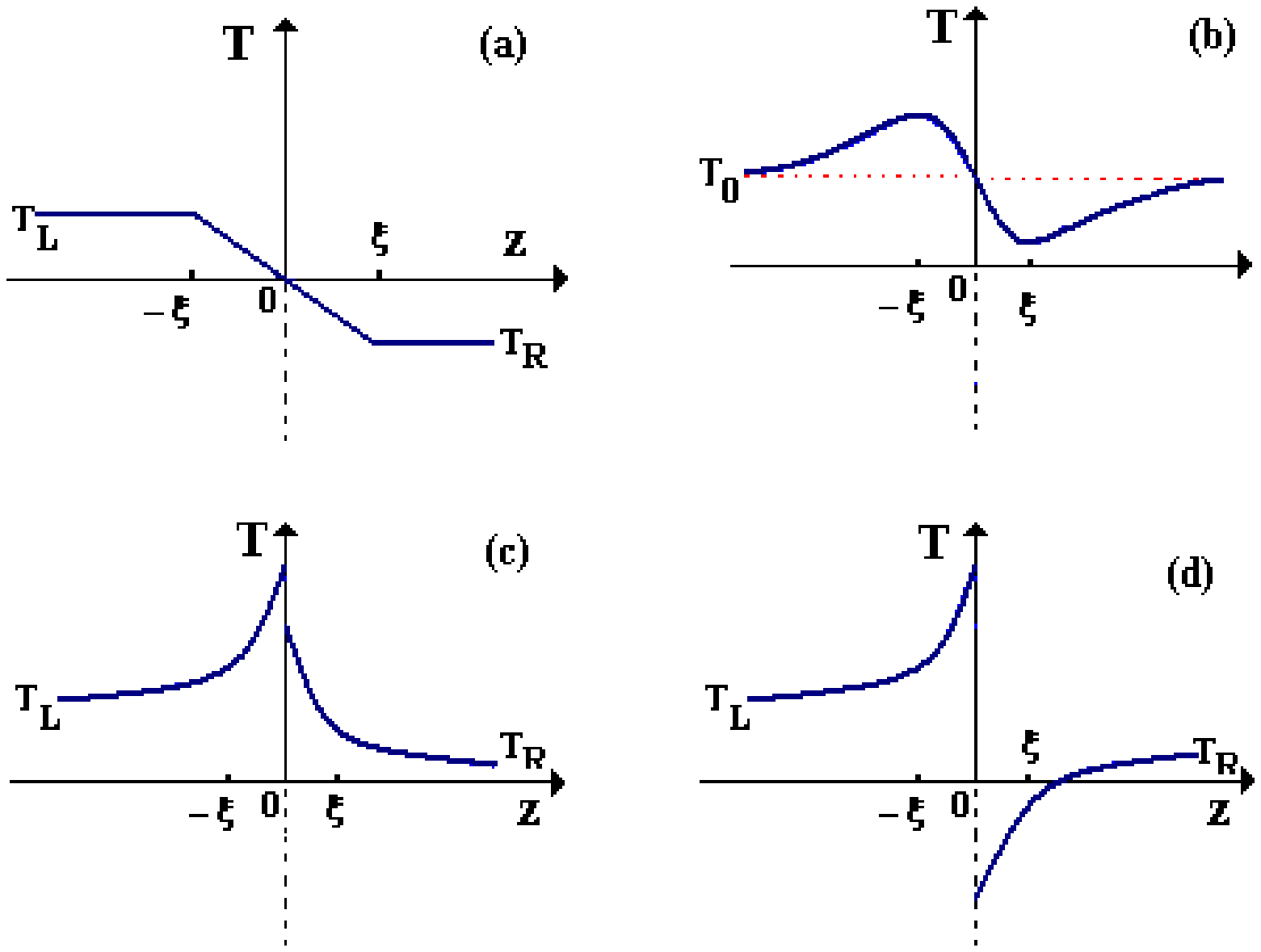} }
\vskip 0.9truecm \caption{Cross section of temperature profile for
different categories. (a) Temperature is continuous on the plate
but its limiting values at $\pm \infty$ are not equal. (b)
Temperature is continuous on the plate and its values at $\pm
\infty$ are equal. (c) and (d) Temperature is not continuous on
the plate.
 \hskip12truecm } \label{oneplateprofile}
\end{figure}

\subsection{Continuous on plate and $T({-\infty})\neq T({\infty})$}

We assume a specific profile for the temperature as [see Fig.
\ref{oneplateprofile}(a)]
\begin{equation}
\begin{array}{l}
T(z)=\left\{
  \begin{array}{l}
    T_L~~~~~~~~~~~~~~~~~~~~~~~~~~~~~~~~~~~~~;~z\leq -\xi, \nonumber\\
    -\frac{1}{2}(T_L-T_R) \frac{z}{\xi}+\frac{1}{2}(T_L+T_R)~~;~-\xi\leq z\leq \xi, \nonumber\\
    T_R~~~~~~~~~~~~~~~~~~~~~~~~~~~~~~~~~~~~~;~\xi\leq z, \nonumber \\
  \end{array}
                   \right.\label{1P-Ta}
\end{array}
\label{T1}
\end{equation}
which exemplifies the general class where the two half-spaces are
kept at different temperatures $T_L$ and $T_R$ and connected by a
crossover domain of size $\xi$. Using the one-plate Green's
function from Eq. (\ref{G-solution}) above, we can calculate the
total stress on the plate as
\begin{equation}
\Pi_{\rm tot}=c_1 \frac{k_{\rm B} (T_L-T_R)}{\xi^2
a},\label{pi-tot-1}
\end{equation}
where $a$ is a microscopic length scale, and $c_1$ is a
nonuniversal numerical constant of order unity. The above result
can be rewritten as
\begin{equation}
\Pi_{\rm tot}\sim-\frac{k_{\rm B}}{\xi a} \; \nabla
T,\label{pi-tot-nabla-1}
\end{equation}
where $\nabla T=\frac{T_R-T_L}{2\xi}$, leading to a Soret
coefficient \cite{definition}
\begin{equation}
S_T \sim \frac{A}{\xi a}
\left(\frac{T_L+T_R}{2}\right)^{-1}.\label{ST1}
\end{equation}
Note that the length scale $\xi$ appears in an anomalous way, in
that it is not represented entirely in the form of $\nabla T$.

\subsection{Continuous on the plate and $T({-\infty})=T({\infty})$}

As another example, which represents the case when the temperature
is continuous on the plate and has equal asymptotic values on the
far left and right sides of the plate, we choose [see
Fig.~\ref{oneplateprofile}(b)]
\begin{equation}
\begin{array}{l}
T(z)=\left\{
  \begin{array}{l}
    T_0-\delta T_0 (e^{-{z}/{\xi}}-e^{-m {z}/{\xi}})~~~~~~~~~~~~~~;~z>0, \nonumber \\
    T_0+\delta T_0 (e^{{z}/{\xi}}-e^{m
    {z}/{\xi}})~~~~~~~~~~~~~~~~~~;~z<0,
    \nonumber \\
  \end{array}
                   \right.
\end{array}
\label{T2}
\end{equation}
for any arbitrary value of $m$. The above temperature profile
yields
\begin{equation}
\Pi_{\rm tot}=c_2 (m-1) \frac{k_{\rm B} \delta T_0}{\xi
a^2},\label{pi-tot-2}
\end{equation}
where $c_2$ is a numerical prefactor of order one. The above
result can be recast in the form
\begin{equation}
\Pi_{\rm tot}\sim - \frac{k_{\rm B}}{a^2} \nabla
T,\label{pi-tot-nabla-2}
\end{equation}
where $\nabla T=(1-m) {\delta T_0}/{\xi}$, from which a Soret
coefficient can be deduced as
\begin{equation}
S_T \sim \frac{A}{a^2 T_0}.\label{ST2}
\end{equation}
Note that in this case the length scale $\xi$ appears only through
the combination $\nabla T$.

\subsection{discontinuous on the plate}   \label{subsec:dis}

Finally, we consider a most general class of temperature profiles
as shown in Figs.~\ref{oneplateprofile}(c) and (d), using the
following example
\begin{equation}
\begin{array}{l}
T(z)=\left\{
  \begin{array}{l}
    T_{R}+\delta T_{+}e^{-{z}/{\xi}}~~~~~~~~~~~~~~~~;~z>0, \nonumber \\
    T_{L}+\delta T_{-}e^{{z}/{\xi}}~~~~~~~~~~~~~~~~~;~z<0, \nonumber \\
  \end{array}
                   \right.
\end{array}
\label{T3}
\end{equation}
Due to the discontinuity of temperature at the boundary, the plate
experiences the strongest form of force that reads
\begin{equation}
\Pi_{\rm tot} \sim -\frac{k_{\rm B} \delta
T}{a^3},\label{pi-tot-3}
\end{equation}
to the leading order, where $\delta T=(T_{R}+\delta
T_{+})-(T_{L}+\delta T_{-})$ is the local temperature difference
across the plate. Note that due to the singular form of the
temperature profile, it is not possible to define a Soret
coefficient in this case.

\section{Two-Plate Geometry} \label{sec:twoplate}

Here we consider the case of two parallel plates immersed in a
nonuniform temperature profile. We commence by the equilibrium
Casimir problem and then go ahead to generalize it to the
nonequilibrium cases. We focus on two different types of behaviors
corresponding to where the temperature profile is uniform or
nonuniform in the exterior regions.

\begin{figure}
\centerline{ \epsfxsize=6truecm \epsffile{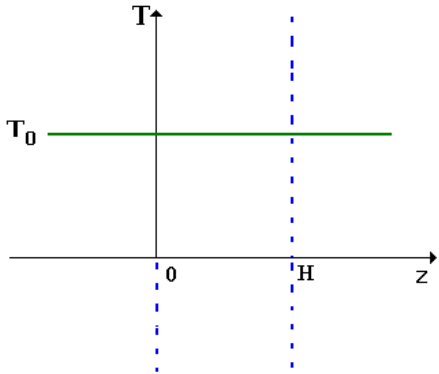} }
\vskip 0.9truecm \caption{Cross section of the temperature profile
for the uniform temperature case. The two plates are located at
$z=0$ and $z=H$.
 \hskip12truecm } \label{twoplateuniform}
\end{figure}

\subsection{Equilibrium Casimir Force}

Here we derive the Fluctuation--induced force on the plates for
the case when temperature profile is uniform everywhere, namely
$T({\bf r},t)=T_0$ as shown in Fig. \ref{twoplateuniform}. The
stress on the plate located at $z=0$ has two contributions coming
from right and left sides. The contribution from the right is
given as
\begin{eqnarray}
\Pi^+&=&-k_{\rm B} T_0 \int\frac{d^2{\bf
q}d\omega}{(2\pi)^3} \left[\frac{\gamma}{K (Q_{+}+Q_{-})} \right. \nonumber\\
&&\left.- \frac{Q_{+}[\coth(Q_{+} H)-1]-Q_{-} [\coth ( Q_{-} H)
-1]}{2i\omega}\right]. \nonumber\\
\label{stress}
\end{eqnarray}
The above stress has two terms; the first term is divergent and
the second one is finite. The contribution from the left side
reads
\begin{equation}
\Pi^-=-\frac{\gamma k_{\rm B} T_0}{K} \int\frac{d^2{\bf
q}d\omega}{(2\pi)^3} \frac{1}{Q_{+}+Q_{-}},\label{stress-left}
\end{equation}
which exactly cancels the divergent part of Eq. (\ref{stress}), so
that the total pressure on this plate is rendered finite as
\begin{equation}
\Pi_{\rm tot}=\frac{\zeta(3)}{8\pi} \frac{k_{\rm B}
T_0}{H^3}.\label{eq-Cas}
\end{equation}
Note that the force is universal, and attractive (as it tends to
push the plate at $z=0$ to the right). We can easily check that
the other plate experiences the same amount of pressure in the
reversed direction, and hence, the Newton's third law applies. As
we will see below, this is a direct consequence of the fact that
this force is mediated by equilibrium fluctuations.

\begin{figure}
\centerline{ \epsfxsize=6truecm \epsffile{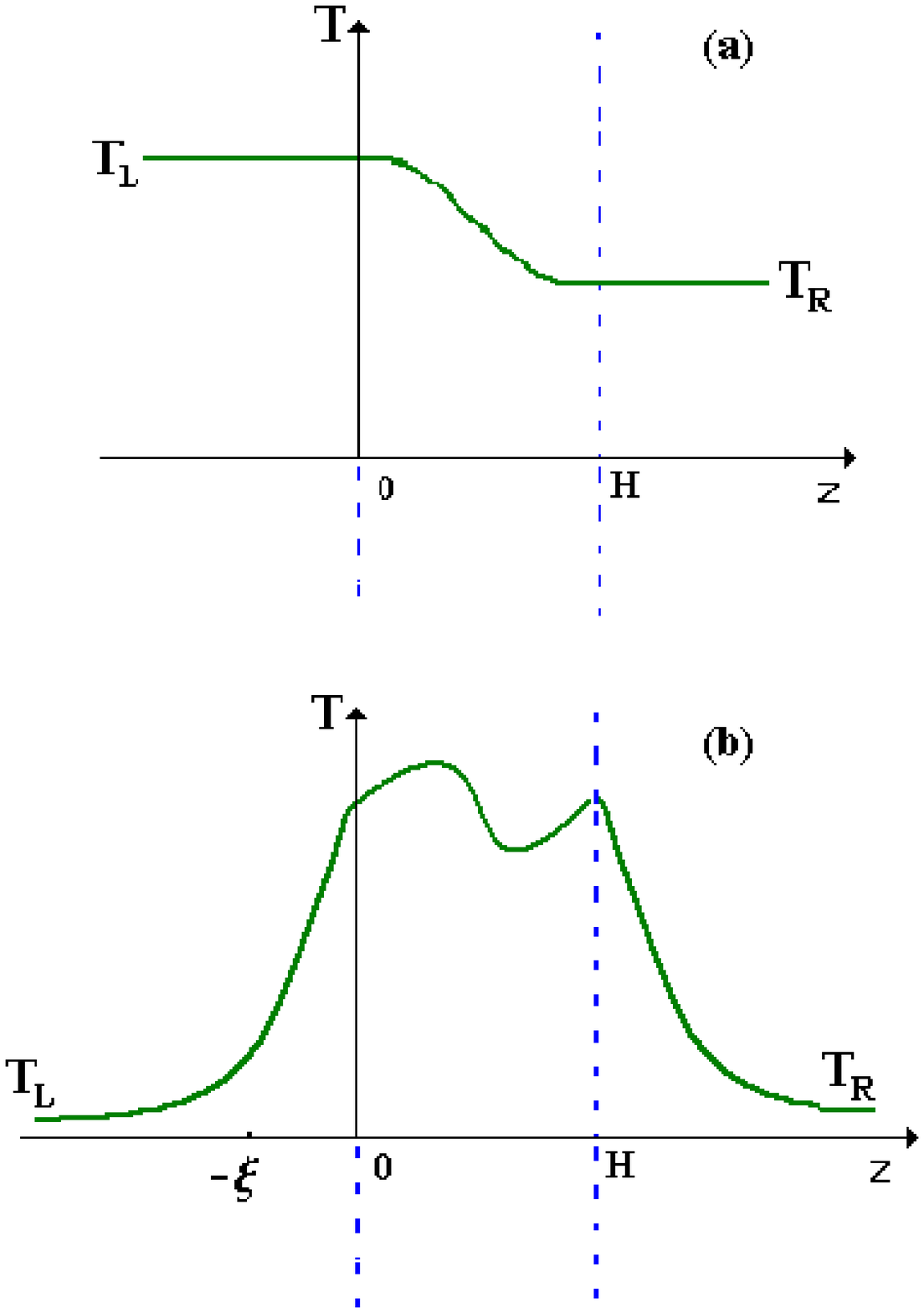} }
\vskip 0.9truecm \caption{Cross section of the temperature profile
for the non-uniform temperature case. The two plates are located
at $z=0$ and $z=H$. (a) Temperature is uniform in the exterior.
(b) Temperature is nonuniform in the exterior. In both cases the
$z$-dependence of the temperature profile the interior is
arbitrary.
 \hskip12truecm } \label{twoplatenonuniform}
\end{figure}

The cancellation of the divergent parts in Eqs. (\ref{stress}) and
(\ref{stress-left}) is directly related to the assumption that
temperature is continuous on the plates. If we considered the case
where temperature is uniform but its value is, say, $T_0$ in the
region M and $T_0'$ in the region L, then in addition to the above
Casimir force we would obtain a divergent part proportional to
$(T_0-T_0')$, similar to the single-plate forces discussed in Sec.
\ref{subsec:dis} above.

\subsection{Temperature profile independent of $z$ in the exterior}

To extend the above calculations to the nonequilibrium situations,
we first consider the case where temperature is independent of $z$
in the exterior regions (R and L) but it is $z$-dependent in the
interior region (M), as shown in Fig. \ref{twoplatenonuniform}(a).
Moreover, we also maintain the possibility of the dependence of
temperature on time, as well as the in-plane coordinates ${\bf
r}$, and assume that the temperature is continuous on the plates
to avoid the prescribed singularities.

We can calculate the two contributions to the local stress on the
plate located at $z=0$ as
\begin{eqnarray}
\Pi^{+}({\bf r},t)&=&-\frac{\gamma k_B}{8K} \int\frac{d^2{\bf
q}d\omega}{(2\pi)^3} \int_{0}^{H} d z' T({\bf
q},L-z',\omega)\nonumber \\
&\times& \int\frac{d^2{\bf q}'d\omega
'}{(2\pi)^3}\frac{\sinh(P_{+} z') \sinh(P_{-}z')}{\sinh (P_{+}H)
\sinh (P_{-} H)}\; e^{i({\bf q} \cdot {\bf r}-\omega
t)},\nonumber\\
\label{NUS}
\end{eqnarray}
and
\begin{eqnarray}
\Pi^{-}&=&-\frac{\gamma k_B}{8K} \int\frac{d^2{\bf
q}d\omega}{(2\pi)^3}  \; T({\bf
q},0,\omega) \nonumber \\
&& \; \; \times \; \int\frac{d^2{\bf q}'d\omega
'}{(2\pi)^3}\frac{1}{P_{+}+P_{-}} \; e^{i({\bf q} \cdot {\bf
r}-\omega t)},\label{left-NUS}
\end{eqnarray}
where $P_{\pm}=\sqrt{\frac{1}{4}({\bf q}\pm{\bf
q}')^2+i\frac{\gamma}{2 K}(\omega\pm\omega ')}$, and we have used
the Fourier transform of temperature in time and in the in-plane
coordinates. Due to the continuity of temperature, the divergent
parts of the two contributions cancel each other as expected, and
the net stress on the plate can be written as (see the Appendix)
\begin{eqnarray}
\Pi({\bf r},t)&=&\frac{\zeta(3)}{8 \pi}\frac{k_{\rm B} T({\bf
r},0,t)}{H^3}-\frac{k_{\rm B}}{8\pi H^3} \int\frac{d^2{\bf
q}d\omega}{(2\pi)^3} \; T({\bf q},0,\omega) \nonumber \\
&& \;\; \; \times \; f\left(\frac{1}{4} q^2 H^2+i\frac{\gamma}{2
K} \omega H^2\right) \; e^{i({\bf q} \cdot {\bf r}-\omega
t)},\label{Pi-tot-twoplate1}
\end{eqnarray}
where
\begin{eqnarray}
f(u)&=& \zeta(3)-\frac{4}{3} u^{3/2}+2 u \; \ln(1-e^{2 \sqrt{u}}) \nonumber \\
&& +2 \sqrt{u} \; {\rm Li}_2(e^{2 \sqrt{u}})-{\rm Li}_3(e^{2
\sqrt{u}}),\label{f(u)1}
\end{eqnarray}
with ${\rm Li}_n(z)=\sum_{k=1}^{\infty} z^k/k^n$ being the
polylogarithm function. The function $f(u)$ is analytic in the
upper-half complex plane is asymptotes to $\zeta(3)$ everywhere on
the semicircle at infinity. For small values of $u$, however, it
has a power series expansion as: $f(u) \simeq u-\frac{2}{3}
u^{3/2}+\frac{1}{6} u^2+O(u^{5/2})$. An interesting feature of the
above result is that the stress on the plate is not sensitive to
the $z$-dependence of the temperature profile in the interior
region.

In the case of uniform temperature Eq. (\ref{Pi-tot-twoplate1})
recovers the equilibrium Casimir force [Eq. (\ref{eq-Cas})]
because temperature will only have the zero frequency and
wavevector component in Fourier space, and the correction term
will vanish due to the factor of $f(0)=0$. When temperature is
modulated in time or in-plane space directions, the plate will
experience local and instantaneous Casimir-like pressure as given
by the first term in Eq. (\ref{Pi-tot-twoplate1}), while the
second term yields corrections as
\begin{eqnarray}
\Pi({\bf r},t)&=&\frac{\zeta(3)}{8\pi} \frac{k_{\rm B} T({\bf
r},0,t)}{H^3} \nonumber\\
&+&\frac{k_{\rm B}}{16 \pi H} \left(\frac{1}{2}
\partial_{\bf r}^2 +\frac{\gamma}{K} \partial_t \right)T({\bf
r}_\parallel,0,t),\label{Pi-tot-twoplate2}
\end{eqnarray}
to the leading order. Note that the above result is an expansion
in powers of the ratio $\frac{H}{\lambda}$, where $\lambda$ is the
typical wavelength set by the variations of the temperature in
space or time, and is valid in the limit $\frac{H}{\lambda} \ll
1$.

In the opposite limit of $\frac{H}{\lambda} \ll 1$, the
fluctuation--induced interactions are screened by the length scale
$\lambda$. This is reminiscent of the screening of the electric
field for periodic charge distributions, where the screening
length is also set by the period of the charge distribution. The
screening manifests itself in the result of Eq.
(\ref{Pi-tot-twoplate1}), where for large values of the wavevector
or frequency the function $f(u)$ tends to $\zeta(3)$, thus leading
to a cancellation of the first term.

The above calculations can be repeated for the second plate to
yield similar expressions for the stress, except that the value of
the temperature will be replaced by its local values in the
neighborhood of the second plate. In the general case when the two
values are different, the two plates will experience different
stresses and thus the third law of Newton will be violated. This
effect is due to the nonequilibrium nature of the fluctuations,
and has also been reported recently in the theory of depletion
forces \cite{depletion}.

When temperature varies only with $z$, we can simplify the
expressions for the stress on the left plate as
\begin{equation}
\Pi_{\rm tot}(0)=\frac{\zeta(3)}{8\pi} \frac{k_{\rm B}
T(0)}{H^3},\label{Casi1}
\end{equation}
and correspondingly for the right plate as
\begin{equation}
\Pi_{\rm tot}(H)=-\frac{\zeta(3)}{8\pi} \frac{k_{\rm B}
T(H)}{H^3}.\label{Casi2}
\end{equation}
Considering the fact that Newton's third law is violated, one can
extract the pressure exerted on the system of the two plates as
\begin{equation}
\Pi_{\rm tot}^{2p}=-\frac{\zeta(3)}{8\pi} \frac{k_{\rm B}
[T(H)-T(0)]}{H^3}.\label{Casi10}
\end{equation}
Note that the above result is reminiscent of Eq. (\ref{pi-tot-3}),
provided that we replace the microscopic length scale $a$ by $H$,
which is the effective thickness of the compound object.

\subsection{Nonuniform temperature profile in the exterior}

The fact that the fluctuation--induced interactions are
independent of the interior temperature inhomogeneity in the $z$
direction is a general result and does not depend on whether or
not the exterior temperature is $z$-dependent. However, the form
of the force that is acting on each plate depends on the
neighboring exterior temperature profile.

To study this class of temperature profiles, we consider the
simple case where the temperature profile is independent of time
and the in-plane space coordinates. We assume the following
profile for the temperature
\begin{equation}
\begin{array}{l}
T(z)=\left\{
  \begin{array}{l}
    T_L+(T-T_L)\;e^{{z}/{\xi}}~~~~~~~~~~~;~z\leq 0, \nonumber\\
    {\rm arbitrary}~~~~~~~~~~~~~~~~~~~~~~~~;~0 < z < H, \nonumber \\
    T_R+(T-T_R)\;e^{-(z-H)/{\xi}}~~~;~z\geq H, \nonumber \\
  \end{array}
                   \right.
\end{array}
\end{equation}
as a typical example for the class of profiles depicted in Fig.
\ref{twoplatenonuniform}(b). Note that we also assume that
temperature is continuous on each plate. We can then calculate the
stress on the left plate as
\begin{equation}
\Pi_{\rm tot}(0)=\frac{\zeta(3)}{8\pi} \frac{k_{\rm B}
T}{H^3}+c_3\frac{k_{\rm B}(T_L-T)}{\xi a^2},\label{Casi3}
\end{equation}
and correspondingly on the right plate as
\begin{equation}
\Pi_{\rm tot}(H)=-\frac{\zeta(3)}{8\pi} \frac{k_{\rm B}
T}{H^3}+c_3\frac{k_{\rm B}(T-T_R)}{\xi a^2}.\label{Casi4}
\end{equation}
We can extract the pressure exerted on the system of the two
plates as
\begin{equation}
\Pi_{\rm tot}^{2p}=c_3 \frac{k_{\rm B}(T_L-T_R)}{\xi
a^2},\label{Casi5}
\end{equation}
which is reminiscent of Eq. (\ref{pi-tot-3}), and thus yields a
Soret coefficient of the form given in Eq. (\ref{ST2}) above.

\section{Concluding remarks}    \label{sec:conc}

In conclusion, we have shown that external objects located in a
nonuniform temperature profile experience nonuniversal
fluctuation--induced forces that tend to move them towards the
colder region, which is a manifestation of the so-called Soret
effect. This effect is examined in various temperature profiles,
and it is shown that the behavior of the fluctuation--induced
forces strongly depends on: (1) whether the temperature is
continuous across the object, (2) the asymptotic values of the
temperature profile, and (3) temporal or spatial variations of
temperature. In the case of two external objects immersed in a
medium undergoing nonequilibrium thermal fluctuations, it is shown
that the third law of Newton is violated in that the two objects
experience different forces depending on where they are located.

\acknowledgements

It is our pleasure to acknowledge stimulating discussions with A.
Ajdari, D. Bartolo, and N. Rasuli.

\appendix

\section{Integration over a Kernel}

\begin{figure}

\centerline{ \epsfxsize=6truecm \epsffile{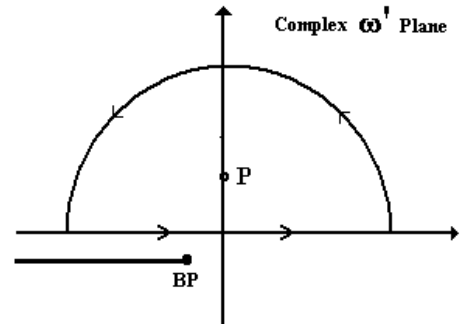} }
\vskip0.9truecm \caption{Complex $\omega '$ plane and the contour
that is used in the integration. The point $P=i K ({\bf q}.{\bf
q}')/\gamma$ is the pole and
$BP=\omega-\frac{i}{2}\frac{K}{\omega}({\bf q}-{\bf q}')^2$ is the
branch point corresponding to the branch cut.
 \hskip12truecm } \label{contour}
\end{figure}

In this Appendix, we outline some details of the calculations
involved in Eqs. (\ref{NUS}) and (\ref{left-NUS}) above. We are
dealing with the following integral:
\begin{eqnarray}
U({\bf q},{\bf
q}',\omega,z')&=&\int_{-\infty}^{\infty}\frac{d\omega '}{2\pi}
\frac{\sinh(P_{+} z') \sinh(P_{-}z')}{\sinh (P_{+}H) \sinh (P_{-}
H)}, \label{kernel1}
\end{eqnarray}
with $P_{\pm}$ being defined below Eq. (\ref{left-NUS}). The
integrand has two branch points, and their corresponding branch
cuts, in the complex $\omega '$ plane. Now, by closing the
integration contour around these cuts, we can directly check that
the integral vanishes for $z'<H$, while it is divergent for the
limiting case of $z'=H$. This behaviour is similar to a
delta-function centered at $z=H$. We can then make the assignment:
$U({\bf q},{\bf q}',\omega,z')=u({\bf q},{\bf q}',\omega) \;
\delta(z'-H)$ and try to find the function $u$. Note that if $u$
is found to be a well-behaved finite function, our assignment will
become rigorous. Integrating both sides of the above equation with
respect to $z'$ yields:
\begin{equation}
u({\bf q},{\bf q}',\omega)=\frac{2i K}{\gamma}\int\frac{d\omega
'}{2\pi} \frac{P_{-} \coth (P_{-} H)}{\omega'-i K ({\bf q}.{\bf
q}')/\gamma}.\label{u-1}
\end{equation}
The integrand of Eq. (\ref{u-1}) has a pole and a branch cut as
shown in Fig.~\ref{contour}. By closing the integration contour in
the upper-half plane we obtain
\begin{eqnarray}
u({\bf q},{\bf
q}',\omega)&=&-\frac{2K}{\gamma}\sqrt{\frac{1}{4}(q^2+q'^2)+i\frac{\gamma}
{2K}\omega} \nonumber
\\
&& \times \; \coth\sqrt{\frac{1}{4}(q^2+q'^2)+i\frac{\gamma}
{2K}\omega},
\end{eqnarray}
which completes the frequency integration.

\end{multicols}


\begin{references}
\bibitem{definition}
The Soret coefficient is defined as follows: the flux of particles
$J$ is related to the temperature gradient via an Onsager relation
$J=-c \; D_T \; \nabla T$, where $c$ is the (particle)
concentration and $D_T$ is the so-called {\it thermal diffusion
coefficient} \cite{soret}. The Soret coefficient is then defined
as $S_T=D_T/D$ where $D$ is the self-diffusion coefficient. Using
the relation $J=c v=c \mu F$ where $\mu$ is the mobility of the
plate and $F=\Pi A$ is the overall force exerted on it, one can
extract a relation between the stress on the plate and the
temperature gradient in terms of the Soret coefficient as:
$\Pi=-\frac{k_{\rm B} T}{A} \; S_T \; \nabla T$, in which the
Einstein's relation $D=\mu \; k_{\rm B} T$ is used.
\end{references}
\end{document}